\documentclass[twocolumn]{autart}  

\usepackage{mypreamble}

\usepackage[T1]{fontenc}

\usepackage[skip=0pt, indent=15pt]{parskip}

\newlength{\eqlen} 
\newlength{\blocklen} 
\begin{document}

\begin{frontmatter}

\title{Reset Controller Analysis and Design for Unstable Linear Plants using Scaled Relative Graphs} 


\author[tue]{Julius P.J. Krebbekx}\ead{j.p.j.krebbekx@tue.nl},    
\author[tue,hungary]{Roland T\'oth}\ead{r.toth@tue.nl},
\author[tue]{Amritam Das}\ead{am.das@tue.nl}  

\address[tue]{Control Systems group, Department of Electrical Engineering,  Eindhoven University of Technology, The Netherlands}                                          
\address[hungary]{Systems and Control Lab, HUN-REN Institute for Computer Science and Control, Budapest, Hungary}


\begin{abstract}                          
    In this technical communique, we develop a graphical design procedure for reset controllers for unstable LTI plants based on recent developments on Scaled Relative Graph analysis, yielding an $L_2$-gain performance bound. The stabilizing controller consists of a second order reset element in parallel with a proportional gain. The proposed method goes beyond existing approaches that are limited to stable systems only, providing a well-applicable approach to design problems in practice where the plant is unstable.  
\end{abstract}

\end{frontmatter}


\section{Introduction}

One of the goals of reset controllers, such as the Clegg integrator~\cite{cleggNonlinearIntegratorServomechanisms1958}, is to have integral action while limiting overshoot. Typically, overshoot is even worse for closed-loop systems that include unstable plants, and performance is limited by the Bode sensitivity integral as well. Therefore, it is of interest to pursue the application of reset controllers to stabilizing unstable plants, and improve their closed-loop performance. 

The \emph{Scaled Relative Graph} (SRG), introduced by Ryu et al.~\cite{ryuScaledRelativeGraphs2022}, offers a graphical tool to analyze feedback systems that are interconnections of input-output systems, represented by operators~\cite{chaffeyGraphicalNonlinearSystem2023}. Recently, an SRG bound has been obtained for a \emph{second order reset element} (SORE) \cite{vandeneijndenScaledGraphsReset2024}, allowing for graphical controller design based on the frequency-domain.

A shortcoming of the state-of-the-art SRG tools developed in \cite{chaffeyGraphicalNonlinearSystem2023} is that they can only be used analyze feedback systems with stable elements. In the case of \emph{Linear Time-Invariant} (LTI) operators, the system to be controlled must be already stable\footnote{This means that the plant to be controller can only have poles with negative real part \cite[Chapter 4]{zhouRobustOptimalControl1996}.}, which poses limitations on practical applicability of SRG tools. In our recent work \cite{krebbekxSRGAnalysisLure2024}, we have shown how to extend SRG analysis to unstable LTI operators by combining the Nyquist criterion~\cite{sandbergFrequencyDomainConditionStability1964} and the SRG for LTI operators as developed in \cite{chaffeyGraphicalNonlinearSystem2023}. 

In this technical communique, we will use the results in~\cite{krebbekxSRGAnalysisLure2024} to stabilize an unstable LTI plant with a SORE in parallel with a proportional gain. Moreover, we will show how the SRG framework provides a graphical tool to design the parallel and proportional gains in the stabilizing controller.

This paper is organized as follows. First, we present the necessary preliminaries in Section~\ref{sec:preliminaries}. Second, in Section~\ref{sec:analysis_of_reset_controllers}, we show how to analyze the stability of a plant in feedback with a reset controller. Finally, we present our main result in Section~\ref{sec:design_procedure}, where we develop a graphical design procedure for a stabilizing controller for unstable LTI systems, starting with a stable reset controller as a building block.

\section{Preliminaries}\label{sec:preliminaries}

In this section, we will briefly state the required mathematical preliminaries for the SRG analysis of systems. For a detailed exposition of the SRG properties see~\cite{ryuScaledRelativeGraphs2022}, and for system analysis with SRGs see~\cite{chaffeyGraphicalNonlinearSystem2023,krebbekxSRGAnalysisLure2024}. Since we are interested in reset controllers, which are switching systems, it is natural to only consider \emph{non-incremental} stability. Therefore, we will see that the Scaled Graph (SG), a special case of the SRG, is of particular interest.

\subsection{Notation and Conventions}

Let $\R, \C$ denote the real and complex number fields, respectively, with $\R_{>0} = (0, \infty)$ and $\C_{\mathrm{Re} > 0}= \{ a+ jb \mid \, a \in \R_{>0}, \, b \in \R \}$, where $j$ is the imaginary unit. Let $\C_\infty := \C \cup \{ \infty \}$ denote the extended complex plane. We denote the complex conjugate of $z = a + jb \in \C$ as $\bar{z} = a-jb$. Let $\mathcal{L}$ denote a Hilbert space, equipped with an inner product $\inner{\cdot}{\cdot}_\mathcal{L} : \mathcal{L} \times \mathcal{L} \to \C$ and norm $\norm{x}_\mathcal{L} := \sqrt{\inner{x}{x}_\mathcal{L}}$.  For sets $A, B \subseteq \C$, the sum and product sets are defined as $A+B:= \{ a+b \mid a\in A, b\in B\}$ and $AB:= \{ ab \mid a\in A, b\in B\}$, respectively. The disk in the complex plane is denoted $D_r(x) = \{ z \in \C \mid |z-x| \leq r \}$. Denote $D_{[\alpha, \beta]}$ the disk in $\C$ centered on $\R$ which intersects $\R$ in $[\alpha, \beta]$. The radius of a set $\mathcal{C} \subseteq \C$ is defined by $\rmin(\mathcal{C}) := \inf_{r>0} : \mathcal{C} \subseteq D_r(0)$. The distance between two sets $\mathcal{C}_1,\mathcal{C}_2 \subseteq \C_\infty$ is defined as $\dist(\mathcal{C}_1,\mathcal{C}_2) := \inf_{z_1 \in \mathcal{C}_1, z_2 \in \mathcal{C}_2} |z_1-z_2|$, where \mbox{$|\infty-\infty|:=0$.}

\subsection{Signals, Systems and Stability}

The Hilbert space of interest is $L_2:= \{ f:\R_{\geq 0} \to \R \mid \norm{f}_2 < \infty \}$, where the norm is induced by the inner product $\inner{f}{g}:= \int_\R f(t) g(t) d t$. The extension of $L_2$, see Ref.~\cite{desoerFeedbackSystemsInputoutput1975}, is defined as $\Lte := \{ u : \R_{\geq 0} \to \R \mid \norm{P_T u}_2 < \infty \text{ for all } T \in \R_{\geq 0} \}$, where $(P_T u)(t) = 0$ for all $T>t$, else $(P_T u)(t)=u(t)$. The extension is particularly useful since it includes periodic signals, which are otherwise excluded from $L_2$. 

Systems are modeled as operators $R: \Lte \to \Lte$. A system is said to be causal if it satisfies $P_T (Ru) = P_T(R(P_Tu))$, i.e., the output at time $t$ is independent of the signal at times greater than $t$. 

Given an operator $R$ on $L_2$, the $L_2$-gain of the operator as is defined (similar to the notation in~\cite{vanderschaftL2GainPassivityTechniques2017}) as 
\vspace{\eqlen}
\begin{equation}\label{eq:non_incremental_induced_norm}
    \gamma(R) := \sup_{u \in L_2} \frac{\norm{Ru}_2}{\norm{u}_2}.\vspace{\eqlen}
\end{equation}
For causal systems, the $L_2$-gain carries over to  $\Lte$ since $\norm{P_T(Ru)}_2 = \norm{P_T(R(P_Tu))}_2 \leq \norm{R(P_Tu)}_2$ and $P_T u \in L_2$ for all $u \in \Lte$ (see Ref.~\cite{vanderschaftL2GainPassivityTechniques2017}). A causal system $R$ is called $L_2$-stable if $\norm{u}_2 < \infty$ implies $\norm{Ru}_2 < \infty$. \emph{Unless specified otherwise, causality and the property $R(0)=0$ is assumed throughout.}


    
    
    

\subsection{Scaled Relative Graphs}

Let $\mathcal{L}$ be a Hilbert space of signals, and $R : \dom(R) \subseteq \mathcal{L} \to \mathcal{L}$ a relation (a possibly multi-valued map, see~\cite{ryuScaledRelativeGraphs2022}). The angle between $u, y\in \mathcal{U} \subseteq \dom(R)$ is defined as 
\vspace{-1em}
\begin{equation}\label{eq:def_srg_angle}
    \angle(u, y) := \cos^{-1} \frac{\mathrm{Re} \inner{u}{y}}{\norm{u} \norm{y}} \in [0, \pi]. \vspace{-1em}
\end{equation}
Given $u_1, u_2 \in \mathcal{U}$, we define the set of complex numbers
\vspace{\eqlen}
\begin{multline*}
    z_R(u_1, u_2) := \\ \left\{ \frac{\norm{y_1-y_2}}{\norm{u_1-u_2}} e^{\pm j \angle(u_1-u_2, y_1-y_2)} \mid y_1 \in R u_1, y_2 \in R u_2 \right\}. \vspace{-1em}
\end{multline*}
The SRG of $R$ over the set $\mathcal{U}$ is defined as
\vspace{\eqlen}
\begin{equation*}
    \SRG_\mathcal{U} (R) := \bigcup_{u_1, u_2 \in \mathcal{U}} z_R(u_1, u_2) \subseteq \C_\infty. \vspace{\eqlen}
\end{equation*}
When $\mathcal{U}=\dom(R)$, we denote $\SRG_{\dom(R)}(R) = \SRG(R)$. The \emph{Scaled Graph} (SG) of an operator $R$ with one input $u^\star \in \mathcal{L}$ fixed and the other in set $\mathcal{U} \subseteq \mathcal{L}$ is defined as
\vspace{\eqlen}
\begin{equation}
    \SG_{\mathcal{U}, u^\star}(R) := \{ z_R(u, u^\star) \mid u \in \mathcal{U} \}. \vspace{\eqlen}
\end{equation}
We use the shorthand $\SG_{\dom(R), u^\star}(R) = \SG_{u^\star}(R)$ when $\mathcal{U}=\dom(R)$. The SG around $u^\star=0$, referred to as the SG of $R$, is of particular interest and is denoted as $\SG(R) := \SG_0(R)$. By definition of the SG, the $L_2$-gain of a system $R$, defined in Eq.~\eqref{eq:non_incremental_induced_norm}, is equal to the radius of the SG of $R$, i.e. $\gamma(R) = \rmin(\SG(R))$. Since we are interested in reset controllers, i.e. switching systems, it is natural to consider non-incremental stability. Therefore, bounding the SG of a feedback interconnection will be the central goal of SRG analysis.


The following proposition describes how elementary operations affect the SG. The result is proven in~\cite[Chapter 4]{ryuScaledRelativeGraphs2022} for the SRG, and trivially carries over to $\SG$ as long as all the operators $R,S$ satisfy $R(0)=S(0)=0$ (since then, $0\in \dom(R)$ and $0\in \ran(R)$ is guaranteed).

\vspace{\blocklen}
\begin{proposition}\label{prop:srg_calculus}
    Let $0 \neq \alpha \in \R$ and let $R, S$ be arbitrary operators on the Hilbert space $\mathcal{L}$. Then, 
    \begin{enumerate}[label=\alph*.]
        \item\label{eq:srg_calculus_alpha} $\SG(\alpha R) = \SG(R \alpha) = \alpha \SG(R)$,
        \item\label{eq:srg_calculus_plus_one} $\SG(I + R) = 1 + \SG(R)$, where $I$ denotes the identity on $\mathcal{L}$,
        \item\label{eq:srg_calculus_inverse} $\SG(R^{-1}) = (\SG(R))^{-1}$, where $R^{-1}$ is defined via the relation~\cite{ryuScaledRelativeGraphs2022} and inversion of $z = re^{j \phi} \in \C$ is defined as the M\"obius inversion $r e^{j \phi} \mapsto (1/r)e^{j \phi}$,
    \end{enumerate}
    If the SGs above contain $\infty$ or are empty sets, the above operations are slightly different, see~\cite{ryuScaledRelativeGraphs2022}. 
\end{proposition}

\subsection{Stability and Performance Analysis using SRGs}

\begin{figure}[tb]
    \centering

    \tikzstyle{block} = [draw, rectangle, 
    minimum height=2em, minimum width=2em]
    \tikzstyle{sum} = [draw, circle, node distance={0.5cm and 0.5cm}]
    \tikzstyle{input} = [coordinate]
    \tikzstyle{output} = [coordinate]
    \tikzstyle{pinstyle} = [pin edge={to-,thin,black}]
    
    \begin{tikzpicture}[auto, node distance = {0.3cm and 0.5cm}]
        \node [input, name=input] {};
        \node [sum, right = of input] (sum) {$ $};
        \node [block, right = of sum] (lti) {$G(s)$};
        \node [coordinate, right = of lti] (z_intersection) {};
        \node [output, right = of z_intersection] (output) {}; 
        \node [block, below = of lti] (static_nl) {$\phi$};
    
        \draw [->] (input) -- node {$r$} (sum);
        \draw [->] (sum) -- node {$e$} (lti);
        \draw [->] (lti) -- node [name=z] {$y$} (output);
        \draw [->] (z) |- (static_nl);
        \draw [->] (static_nl) -| node[pos=0.99] {$-$} (sum);
    \end{tikzpicture}
    
    \caption{Block diagram of a Lur'e system.}
    \label{fig:lure}
\end{figure}
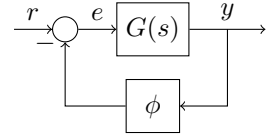

We focus on the Lur'e system in Fig.~\ref{fig:lure} where $G(s)$ is an LTI plant and $\phi$ is a nonlinear operator on $L_{2\mathrm{e}}$. We write $T = (G^{-1} + \phi)^{-1}$ for the closed-loop operator of such a Lur'e system. The Lur'e system $y = Tr$ is called \emph{well-posed} if for each input $r \in L_2$, there exist unique $e, y \in \Lte$ such that $e = r-\phi y$ and $y = G e$. Furthermore, we define the Nyquist diagram as $\operatorname{Nyquist}(G) = \{ G(j \omega)\mid \omega \in \R\}$. Definition~\ref{def:extended_srg} and Theorem~\ref{thm:lti_srg_nyquist_extension} present one of the main results in \cite{krebbekxSRGAnalysisLure2024}, which is a generalization of the circle criterion~\cite{sandbergFrequencyDomainConditionStability1964}.

\vspace{\blocklen}
\begin{definition}\label{def:extended_srg}
    Let $R$ be an LTI operator with $n_\mathrm{p}$ poles $p$ with $\mathrm{Re}(p) >0$. Denote the h-convex hull\footnote{See~\cite{chaffeyGraphicalNonlinearSystem2023} for the definition of the h-convex hull.} of $\operatorname{Nyquist}(R)$ as $\mathcal{G}_R$ and define
    \vspace{\eqlen}
    \begin{equation}\label{eq:set_of_encircled_unstable_points}
        \mathcal{N}_R = \{ z \in \C \mid N_R(z) +n_\mathrm{p} >0 \}, \vspace{\eqlen}
    \end{equation}
    where the winding number $N_R(z) \in \Z$ denotes the amount of clockwise encirclements of $z$ by $\operatorname{Nyquist}(R)$\footnote{The D-contour with inner intedations for poles on the imaginary axis is used when counting the encirclements, analogously to the SISO Nyquist criterion used in the LTI case.}. Define the extended SRG of an LTI operator as 
    \vspace{\eqlen}
    \begin{equation}\label{eq:lti_srg_redefinition}
        \SRG'(R) := \mathcal{G}_R \cup \mathcal{N}_R. \vspace{\eqlen}
    \end{equation}
\end{definition}
\vspace{\blocklen}

\noindent We can now state the generalized circle criterion from~\cite{krebbekxSRGAnalysisLure2024}.

\vspace{\blocklen}
\begin{theorem}\label{thm:lti_srg_nyquist_extension}
    Under the condition that $\gamma(\phi)<\infty$, one of $\SRG(G)^{-1}$ or $\SG(\phi)$ obeys the chord property and there exists some $\kappa \in \R$ such that $\kappa \in \SG(\phi)$ and for all $0\leq \tau_1 \leq \tau_2 \leq 1$, it holds that 
    \vspace{\eqlen}
    \begin{equation}\label{eq:phi_homotopy_assumption}
        \tau_1 \left( \SG(\phi) - \kappa \right) \subseteq \tau_2 \left( \SG(\phi) - \kappa \right). \vspace{\eqlen}
    \end{equation}
    Then, if $(G^{-1}+\tau \phi)^{-1}$ is well-posed for all $\tau \in (0,1]$ and
    \vspace{\eqlen}
    \begin{equation}\label{eq:separation_assumption}
        \dist(\SRG'(G)^{-1}, -\SG(\phi)) \geq r >0 \vspace{\eqlen}
    \end{equation}
    holds, the system in Fig.~\ref{fig:lure} is an $L_2$-stable operator $T : L_2 \to L_2$. Furthermore, the closed-loop operator satisfies $\gamma(T) \leq 1/r$.
\end{theorem}
\vspace{\blocklen}

Condition \eqref{eq:phi_homotopy_assumption} is the same as requiring that $\phi - \kappa$ is \emph{star-shaped} for some $\kappa \in \SG(\phi)$. This property is trivial for convex shapes, and easily verified graphically in most cases. The separation condition \eqref{eq:separation_assumption} is checked graphically, analogously to the circle criterion~\cite{sandbergFrequencyDomainConditionStability1964}, and involves evaluating the winding number in Eq.~\eqref{eq:set_of_encircled_unstable_points} for all $z \in \C$. 

Note that we must \emph{assume} well-posedness of $T_\tau = (G^{-1} + \tau \phi)^{-1}$ for all $\tau \in (0,1]$ to conclude that the system $T:=T_1$ is $L_2$-stable. This well-posedness assumption is used in~\cite{megretskiSystemAnalysisIntegral1997}, and argued to be a mild assumption.

\section{Analysis of Reset Controllers}\label{sec:analysis_of_reset_controllers}

In this section we will use recent results in SRG analysis~\cite{vandeneijndenScaledGraphsReset2024,krebbekxSRGAnalysisLure2024} to analyze the stability and $L_2$-gain performance LTI systems, which may be unstable, in feedback with a reset controller.

\subsection{The Reset Controller}\label{sec:reset_controllers}

This subsection recapitulates the relevant material from~\cite{vandeneijndenScaledGraphsReset2024}, which is an SG bound for a SORE element. We consider continuous-time reset systems $\mathcal{R}$ that can be represented as 
\vspace{\eqlen}
\begin{equation*}
    \mathcal{R} :
    \begin{cases}
        \dot x(t) = A x(t) + B u(t), & \text{ if } (x(t), u(t)) \in \mathcal{F}, \\
        x(t^+) = R_\mathcal{J} x(t), & \text{ if } (x(t), u(t)) \in \mathcal{J}, \\
        y(t) = C x(t) + D u(t),
    \end{cases} \vspace{\eqlen}
\end{equation*}

where $x(t) \in \R^n$ is the state with dimension $n \in \N$, $u(t) \in \R$ the input and $y(t) \in \R$ the output at time $t \in [0, \infty)$. We define $x(t^+) := \lim_{\tau \downarrow t} x(\tau)$. The matrices $A \in \R^{n \times n}, B \in \R^{n \times 1}, C \in \R^{1 \times n}, D \in \R$ define the state-space model and the matrix $R_\mathcal{J} \in \R^{n \times n}$ is the reset map. The sets $\mathcal{F} = \{ \xi \in \R^{n+1} \mid \xi^\top M \xi \geq 0 \}$ and $\mathcal{J} = \{ \xi \in \R^{n+1} \mid \xi^\top M \xi \leq 0 \}$ denote the flow and jump sets, respectively, where $\xi = [x^\top \, u^\top]^\top$ and $M=M^\top \in \R^{(n+1) \times (n+1)}$ is the reset condition matrix.

\vspace{\blocklen}
\begin{example}\label{example:reset-controller}
    Consider the reset controller defined by $R_\mathcal{J} = 0 \in \R^{2 \times 2}$ and 
    \vspace{\eqlen}
    \begin{equation}\label{eq:reset_controller_eqs}
    \begin{split}
        \left[
        \begin{array}{c|c}
        A & B\\
        \hline
        C & D
        \end{array}
        \right] = 
        \left[
        \begin{array}{cc|c}
        -1 & 0 & 1 \\
        1 & -1 & 0 \\
        \hline
        0 & 1 & 0
        \end{array}
        \right], \quad 
        M =  
        \left[
        \begin{array}{cc:c}
        \alpha^2 & 0 & 0 \\
        0 & -1 & 0 \\
        \hdashline
        0 & 0 & 0
        \end{array}
        \right],
    \end{split}
    \end{equation}
    where $\alpha = 0.9$ is chosen, similar to~\cite[Example 1]{vandeneijndenScaledGraphsReset2024}, so the controller resets whenever $\alpha^2 x_1(t)^2 - x_2(t)^2 \leq 0$. 
\end{example}
\vspace{\blocklen}

The SRG bound obtained for this controller is~\cite{vandeneijndenScaledGraphsReset2024} 
\vspace{\eqlen}
\begin{equation}\label{eq:sore_sg_bound}
\begin{split}
    \SG(\mathcal{R}) \subseteq \mathcal{S} &:= \\ \{ z = r e^{\phi}  \mid \phi & \in [- \pi/2, \pi/2), r\in [-0.504, 0.85]\}.
\end{split} 
\end{equation}
\vspace{\eqlen}

Note that in Example~\ref{example:reset-controller}, $\kappa = 0 \in \mathcal{S}$ and that $\tau_1 \mathcal{S} \subseteq \tau_2 \mathcal{S}$ for all $0 \leq \tau_1 \leq \tau_2 \leq 1$, hence $\mathcal{S}$ satisfies condition~\eqref{eq:phi_homotopy_assumption}. Furthermore, since $\mathcal{S}$ consists of two circle halves, it is easy to see that $\mathcal{S}$ satisfies the chord property. From Eq.~\eqref{eq:sore_sg_bound} it is also clear that $\gamma(\mathcal{R}) \leq 0.85$.

\subsection{Analysis with a stable plant}

When controlling a stable plant with a reset controller, the feedback interconnection under consideration is the Lur'e system in Fig.~\ref{fig:lure}, where $G = G_\mathrm{s}$ is the stable plant and $\phi = \mathcal{R}$ is the given reset controller. To demonstrate how the analysis is done using Theorem~\ref{thm:lti_srg_nyquist_extension}, we consider the following example.

\begin{figure}[tb]
    \centering
    \begin{subfigure}[b]{0.4\linewidth}
    \centering
    \tikzstyle{block} = [draw, rectangle, 
    minimum height=2em, minimum width=2em]
    \tikzstyle{sum} = [draw, circle, node distance={0.5cm and 0.5cm}]
    \tikzstyle{input} = [coordinate]
    \tikzstyle{output} = [coordinate]
    \tikzstyle{pinstyle} = [pin edge={to-,thin,black}]
    
    \begin{tikzpicture}[auto, node distance = {0.3cm and 0.5cm}]
        \node [input, name=input] {};
        \node [sum, right = of input] (sum) {$ $};
        \node [block, right = of sum] (lti) {$G_\mathrm{s}(s)$};
        \node [coordinate, right = of lti] (z_intersection) {};
        \node [output, right = of z_intersection] (output) {}; 
        \node [block, below = of lti] (static_nl) {$C_\mathrm{s}$};
    
        \draw [->] (input) -- node {$r$} (sum);
        \draw [->] (sum) -- node {$e$} (lti);
        \draw [->] (lti) -- node [name=z] {$y$} (output);
        \draw [->] (z) |- (static_nl);
        \draw [->] (static_nl) -| node[pos=0.99] {$-$} (sum);
    \end{tikzpicture}
    \caption{Example~\ref{example:stable}.}
    \label{fig:interconnection}
    \end{subfigure}
    \hfill
    \centering
    \begin{subfigure}[b]{0.59\linewidth}
    \centering
    \tikzstyle{block} = [draw, rectangle, 
    minimum height=2em, minimum width=2em]
    \tikzstyle{sum} = [draw, circle, node distance={0.5cm and 0.5cm}]
    \tikzstyle{input} = [coordinate]
    \tikzstyle{output} = [coordinate]
    \tikzstyle{pinstyle} = [pin edge={to-,thin,black}]
    
    \begin{tikzpicture}[auto, node distance = {0.3cm and 0.5cm}]
        \node [input, name=input] {};
        \node [sum, right = of input] (sum) {$ $};
        \node [block, right = of sum] (lti) {$G_\mathrm{u}(s)$};
        \node [coordinate, right = of lti] (z_intersection) {};
        \node [output, right = of z_intersection] (output) {}; 
        \node [block, below = of lti] (static_nl) {$C_\mathrm{u}$};
    
        \draw [->] (input) -- node {$r$} (sum);
        \draw [->] (sum) -- node {$e$} (lti);
        \draw [->] (lti) -- node [name=z] {$y$} (output);
        \draw [->] (z) |- (static_nl);
        \draw [->] (static_nl) -| node[pos=0.99] {$-$} (sum);
    \end{tikzpicture}
    \caption{Example~\ref{example:unstable}.}
    \label{fig:interconnection_unstable}
    \end{subfigure}
    \caption{Block diagrams for the stable and unstable LTI system with the reset controller.}
    \label{fig:interconnections}
\end{figure}
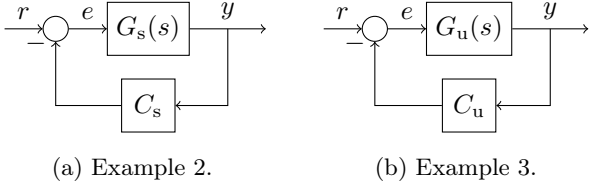

\vspace{\blocklen}
\begin{example}\label{example:stable}
    Consider the system $T_\mathrm{s} = (G_\mathrm{s}^{-1} + C_\mathrm{s})^{-1}$ in Fig.~\ref{fig:interconnection} with $C_\mathrm{s} = \mathcal{R}$ from Example~\ref{example:reset-controller} and
    \vspace{\eqlen}
    \begin{equation}\label{eq:stable}
        G_\mathrm{s}(s) = \frac{14 s + 8}{s^5 + 13 s^4 + 58s^2 + 34 s + 4.2} \vspace{\eqlen}
    \end{equation}
    as is considered in~\cite{vandeneijndenScaledGraphsReset2024}.
\end{example}
\vspace{\blocklen}

\begin{figure}[tb]
    \centering
     \begin{subfigure}[b]{0.46\linewidth}
         \centering
         \includegraphics[width=\linewidth]{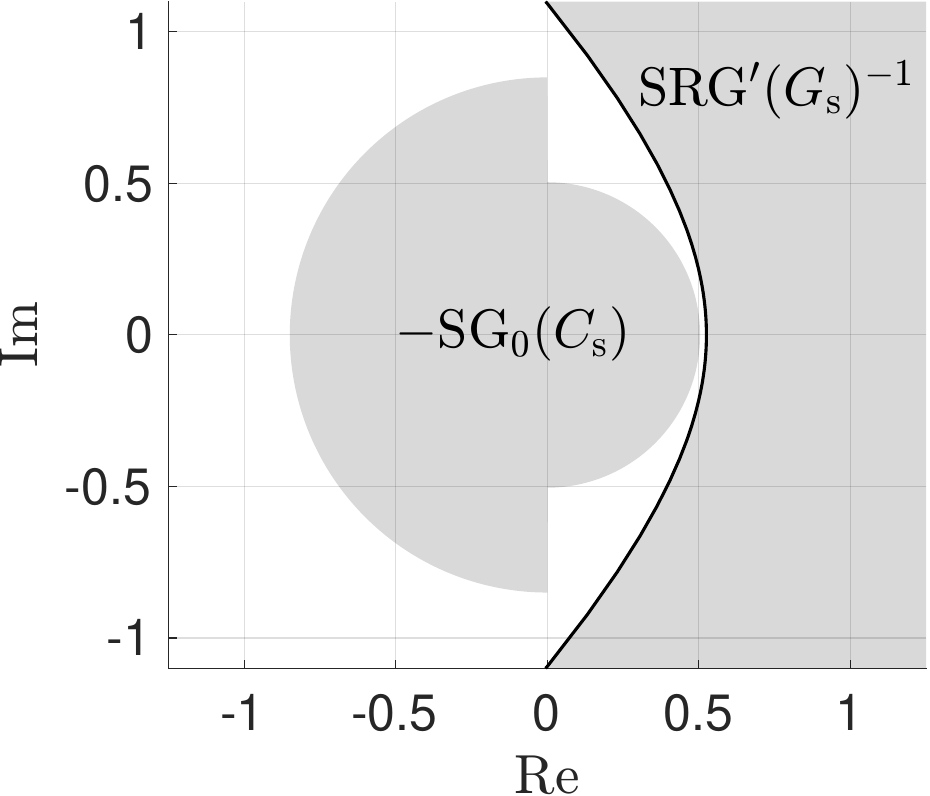}
         \caption{SRG analysis.}
         \label{fig:srg_stable}
     \end{subfigure}
     \hfill
     \centering
     \begin{subfigure}[b]{0.53\linewidth}
         \centering
         \includegraphics[width=\linewidth]{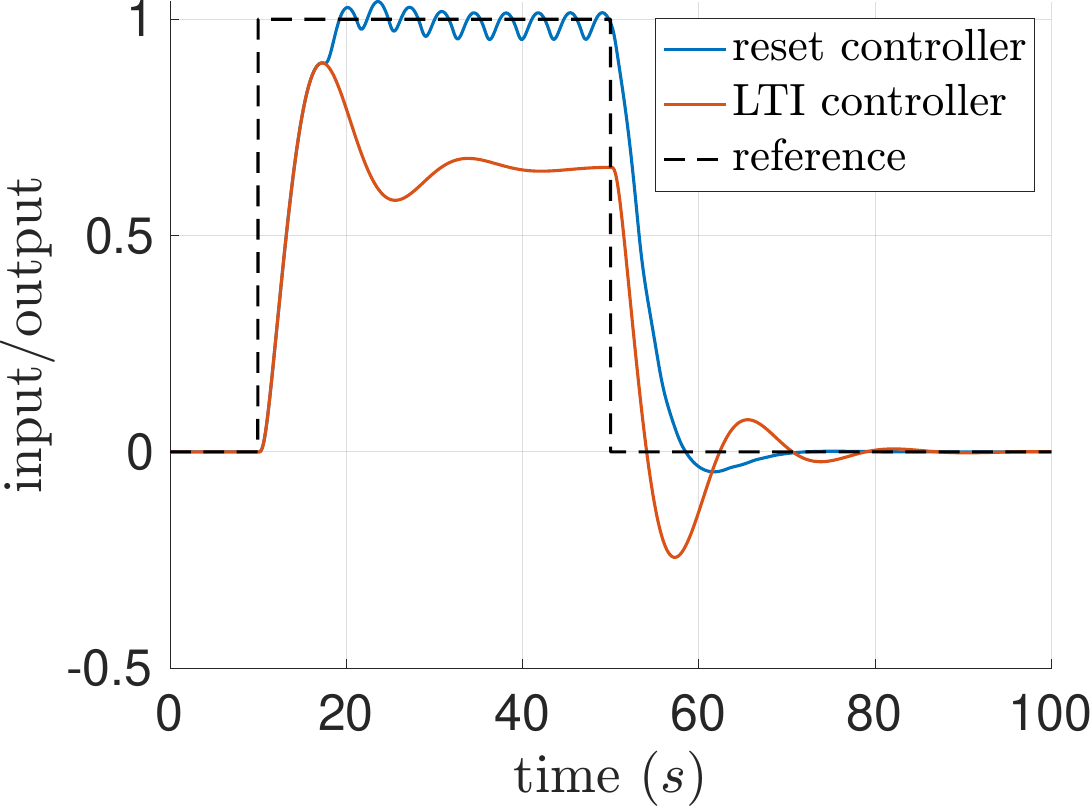}
         \caption{Simulation.}
         \label{fig:sim_stable}
     \end{subfigure}
    \caption{Analysis of Example~\ref{example:stable}.}
    \label{fig:stable}
\end{figure}

In Fig.~\ref{fig:srg_stable}, we show the SRG analysis of Example~\ref{example:stable}. The analysis is performed using Theorem~\ref{thm:lti_srg_nyquist_extension} by taking $G=G_\mathrm{s}$ and $\phi = C_\mathrm{s} = \mathcal{R}$. Firstly, from Section~\ref{sec:reset_controllers}, we know that $\gamma(\mathcal{R}) < \infty$ and $\mathcal{S}$ satisfies the chord property. Secondly, the separation between the graphs $\SRG'(G_\mathrm{s})^{-1}$ and $- \tau \mathcal{S}$ for all $\tau \in (0,1]$, as shown in Fig.~\ref{fig:srg_stable}, proves the separation condition Eq.~\eqref{eq:separation_assumption}. Since we satisfy all conditions for Theorem~\ref{thm:lti_srg_nyquist_extension}, we can conclude stability of the closed-loop (assuming well-posedness of $(G_\mathrm{s}^{-1} + \tau \mathcal{R})^{-1}$ for all $\tau \in [0,1]$), together with an $L_2$-gain bound $\gamma(T_\mathrm{s}) \leq 1/r$, where $r = 0.021$ is the minimal separation of the graphs $\SRG'(G_\mathrm{s})^{-1}$ and $-\mathcal{S}$.

The simulation results in Fig.~\ref{fig:sim_stable} show that the reset controller performs better, in terms of reference tracking, when compared to the LTI controller obtained by removing the reset condition from $\mathcal{R}$. Note that the bound $\gamma(T_\mathrm{s}) \leq 47.6$ is very conservative compared to the response in Fig.~\ref{fig:sim_stable}. This is due to a certain conservatism in the SG bound $\mathcal{S}$ of $\mathcal{R}$, and is discussed further in Section~\ref{sec:reset_controller_bound}. 

\subsection{Analysis with an unstable plant}\label{sec:unstable}

\begin{figure}[tb]
    \centering
     \begin{subfigure}[b]{0.49\linewidth}
         \centering
         \includegraphics[width=\linewidth]{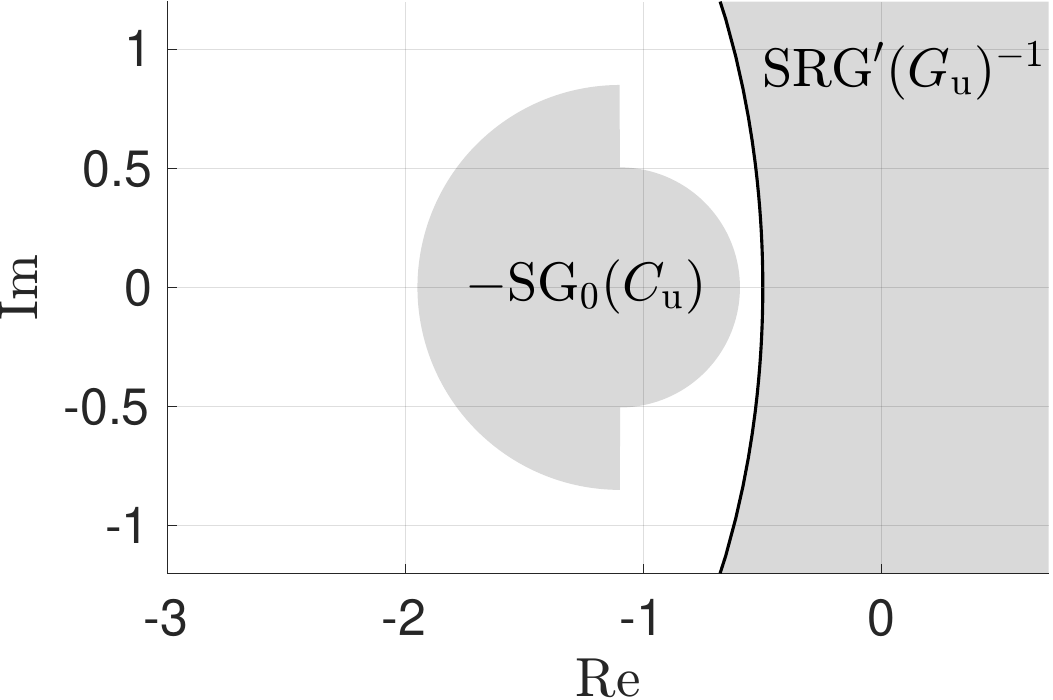}
         \caption{SRG analysis.}
         \label{fig:srg_unstable}
     \end{subfigure}
     \hfill
     \centering
     \begin{subfigure}[b]{0.49\linewidth}
         \centering
         \includegraphics[width=\linewidth]{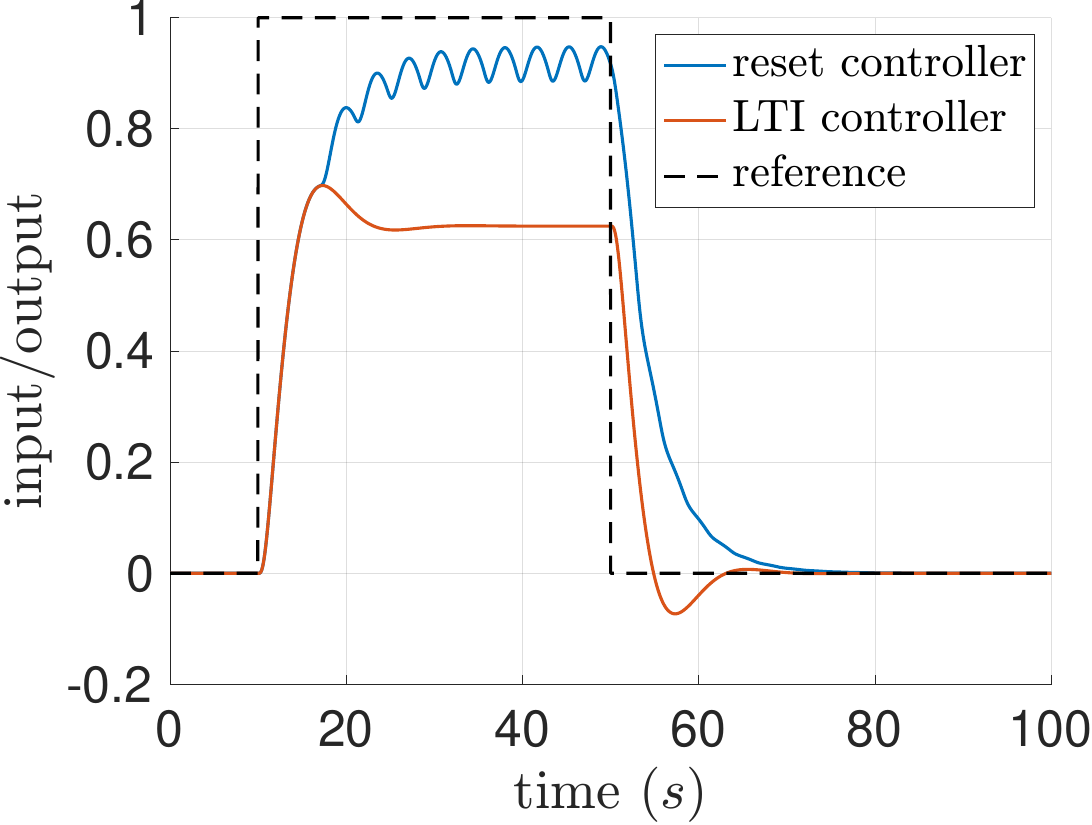}
         \caption{Simulation results.}
         \label{fig:sim_unstable}
     \end{subfigure}
    \caption{Analysis of Example~\ref{example:unstable}.}
    \label{fig:unstable}
\end{figure}

When controlling an unstable plant $G_\mathrm{u}$ with a (reset) controller $C_\mathrm{u}$, the feedback interconnection under consideration is given in Fig.~\ref{fig:interconnection_unstable}. Through the following example, we illustrate the procedure to analyze the closed-loop stability of an unstable plant with given reset controller. 

\vspace{\blocklen}
\begin{example}\label{example:unstable}
    Consider the system $T_{\mathrm{u}} = (G_\mathrm{u}^{-1} + C_{\mathrm{u}})^{-1}$ in Fig.~\ref{fig:interconnection_unstable} with unstable plant
    \vspace{\eqlen}
    \begin{equation}\label{eq:unstable}
        G_\mathrm{u}(s) = \frac{14s + 8}{s^5 + 13s^4 + 58s^3 + 96s^2 + 34s -4}, \vspace{\eqlen}
    \end{equation}
    and reset controller $C_\mathrm{u}=1+1.1\mathcal{R}$, where $\mathcal{R}$ is taken from Example~\ref{example:reset-controller}.
\end{example}
\vspace{\blocklen}

In Example~\ref{example:unstable}, the controller $C_\mathrm{u}$ consists of a SORE $\mathcal{R}$ with gain $1.1$ and a unit proportional term in parallel. In Section~\ref{sec:reset_controllers} it was shown that $\mathcal{R}$ in Example~\ref{example:reset-controller} satisfies the condition in Eq.~\eqref{eq:phi_homotopy_assumption} for $\kappa=0$. Therefore, $C_\mathrm{u}$ in Example~\ref{example:unstable} satisfies Eq.~\eqref{eq:phi_homotopy_assumption} for $\kappa=1$. The SG bound for $\SG(C_\mathrm{u})$ is computed by evaluating $\mathcal{C}_\mathrm{u} = 1 + 1.1 \mathcal{S}$ using Propostion~\ref{prop:srg_calculus}.\ref{eq:srg_calculus_alpha} and \ref{prop:srg_calculus}.\ref{eq:srg_calculus_plus_one}, where $\mathcal{S}$ is taken from Eq.~\eqref{eq:sore_sg_bound} and $\SRG'(G_\mathrm{u})$ is evaluated according to Definition~\ref{def:extended_srg}. The stability analysis is done using Theorem~\ref{thm:lti_srg_nyquist_extension} by taking $G=G_\mathrm{u}, \phi = C_\mathrm{u}$, where Fig.~\ref{fig:srg_unstable} clearly shows a separation between the graphs $\SRG'(G_\mathrm{u})^{-1}$ and $-\mathcal{C_\mathrm{u}}$, which proves \eqref{eq:separation_assumption}. We satisfy all conditions for Theorem~\ref{thm:lti_srg_nyquist_extension}, hence we can conclude stability of the closed-loop (assuming well-posedness of $(G_\mathrm{u}^{-1} + \tau C_\mathrm{u})^{-1}$ for all $\tau \in [0,1]$), together with an $L_2$-gain bound $\gamma(T_\mathrm{u}) \leq 1/r$, where $r = 0.096$ is the minimal separation of the graphs $\SRG'(G_\mathrm{u})^{-1}$ and $-\mathcal{C}_\mathrm{u}$.

The simulation results in Fig.~\ref{fig:sim_unstable} look very similar to those in Fig.~\ref{fig:sim_stable}. In both cases, the reset controller performs better in terms of reference tracking, but creates fast oscillations due to the reset behavior. Again, the $L_2$-gain bound of $\gamma(T_\mathrm{u}) \leq 10.42$ seems conservative when compared to Fig.~\ref{fig:sim_stable}, which is explained by the arguments in Section~\ref{sec:reset_controller_bound}.

\subsubsection{Tighter SG bounds for reset systems}\label{sec:reset_controller_bound}

It was found in simulations that the system in Example~\ref{example:unstable} was still stable even if the right circular part of $-\SG(C_\mathrm{u})$ in Fig.~\ref{fig:srg_unstable} would intersect (for different parameter values) with $\SRG'(G_\mathrm{u})^{-1}$. This can be understood from the recent work by~\cite{grootdeGraphicalTuningParadigm2024}, who showed that the bound \eqref{eq:sore_sg_bound} is conservative. The part that was shown in \cite{grootdeGraphicalTuningParadigm2024} to be an especially conservative over-approximation is the \emph{left} circular part of $\mathcal{S}$ ($r = -0.504$ part of Eq.~\eqref{eq:sore_sg_bound}). This conservatism, however, does not influence the controller design procedure outlined here, and therefore we use Eq.~\eqref{eq:sore_sg_bound} in our examples. 

\section{Graphical controller design procedure}\label{sec:design_procedure}

Based on Example~\ref{example:unstable}, we will now outline a graphical procedure for designing reset controllers for LTI plants. Consider the controller $C(k_\mathrm{p}, k_\mathrm{r}) = k_\mathrm{p} + k_\mathrm{r} \mathcal{R}$, where $k_\mathrm{p}, k_\mathrm{r} \in \R$ are the parallel and reset gains, respectively. Given an LTI plant $G$, the controller design task is to find gains $k_\mathrm{p}, k_\mathrm{r}$ such that the closed-loop $T= (G^{-1} + C(k_\mathrm{p}, k_\mathrm{r}))^{-1}$ is stable and obeys $\gamma(T) \leq \hat{\gamma}$ for some predefined $\hat{\gamma} > 0$. 

We will show that Theorem~\ref{thm:lti_srg_nyquist_extension} gives a natural \emph{graphical} design procedure to find $k_\mathrm{p}, k_\mathrm{r}$. Using Propostion~\ref{prop:srg_calculus}.\ref{eq:srg_calculus_alpha} and \ref{prop:srg_calculus}.\ref{eq:srg_calculus_plus_one}, one obtains the SG bound
\vspace{\eqlen}
\begin{equation}\label{eq:sg_bound_param_controller}
    \SG(C(k_\mathrm{p}, k_\mathrm{r})) \subseteq k_\mathrm{p} + k_\mathrm{r} \mathcal{S}. \vspace{\eqlen}
\end{equation}
Note that $k_\mathrm{p} + k_\mathrm{r} \mathcal{S}$ is inflatable (from $k_\mathrm{p}$) and obeys the chord property (see Section~\ref{sec:reset_controllers}). The following steps describe the design procedure.
\begin{enumerate}
    \item Compute $\SRG'(G)$, and plot $\SRG'(G)^{-1}$.
    \item Choose $k_\mathrm{p}, k_\mathrm{r}$ such that $-k_\mathrm{p} - k_\mathrm{r} \mathcal{S}$ does not intersect $\SRG'(G)^{-1}$.
    \item Tune the parameters $k_\mathrm{p}, k_\mathrm{r}$ such that 
    \vspace{\eqlen}
    \begin{equation}\label{eq:reset_design_dist}
        \operatorname{dist}(\SRG'(G)^{-1}, -k_\mathrm{p} - k_\mathrm{r} \mathcal{S}) \geq 1/\hat{\gamma}. \vspace{\eqlen}
    \end{equation}
\end{enumerate}
Steps (2) and (3) in the procedure rely entirely on \emph{graphical} controller tuning using the SG. We note that when focusing purely on $L_2$-gain, it is always advantageous to choose $k_\mathrm{r}=0$ to maximize the separation in Eq.~\eqref{eq:reset_design_dist}. However, $L_2$-gain does not capture all performance aspects, and it is well-known~\cite{cleggNonlinearIntegratorServomechanisms1958} that adding reset behavior to an LTI controller can increase the performance (e.g. go beyond the Bode sensitivity integral~\cite{aangenentPerformanceAnalysisReset2010}). Therefore, one can choose to fix $k_\mathrm{r}$ in the design procedure. Alternatively, one can maximize $|k_\mathrm{r}|$ while satisfying Eq.~\eqref{eq:reset_design_dist}, in the same spirit as \cite{astromDesignPIControllers1998}. We conclude by demonstrating one design procedure.

\begin{figure}[tb]
    \centering
     \begin{subfigure}[b]{0.49\linewidth}
         \centering
         \includegraphics[width=\linewidth]{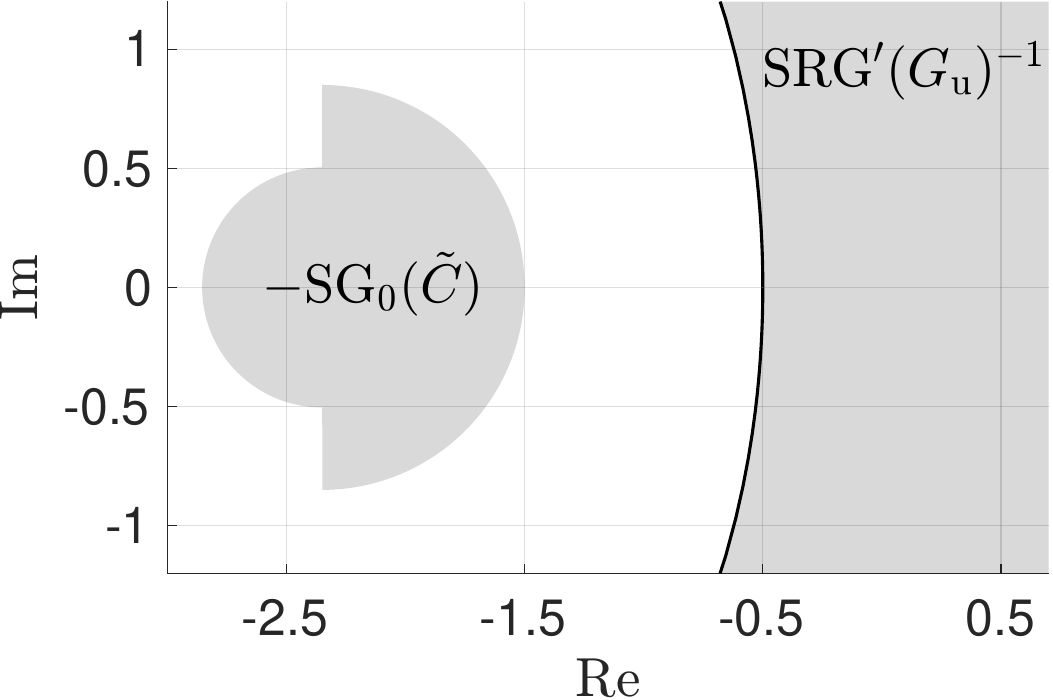}
         \caption{SRG analysis.}
         \label{fig:srg_example_procedure}
     \end{subfigure}
     \hfill
     \centering
     \begin{subfigure}[b]{0.49\linewidth}
         \centering
         \includegraphics[width=\linewidth]{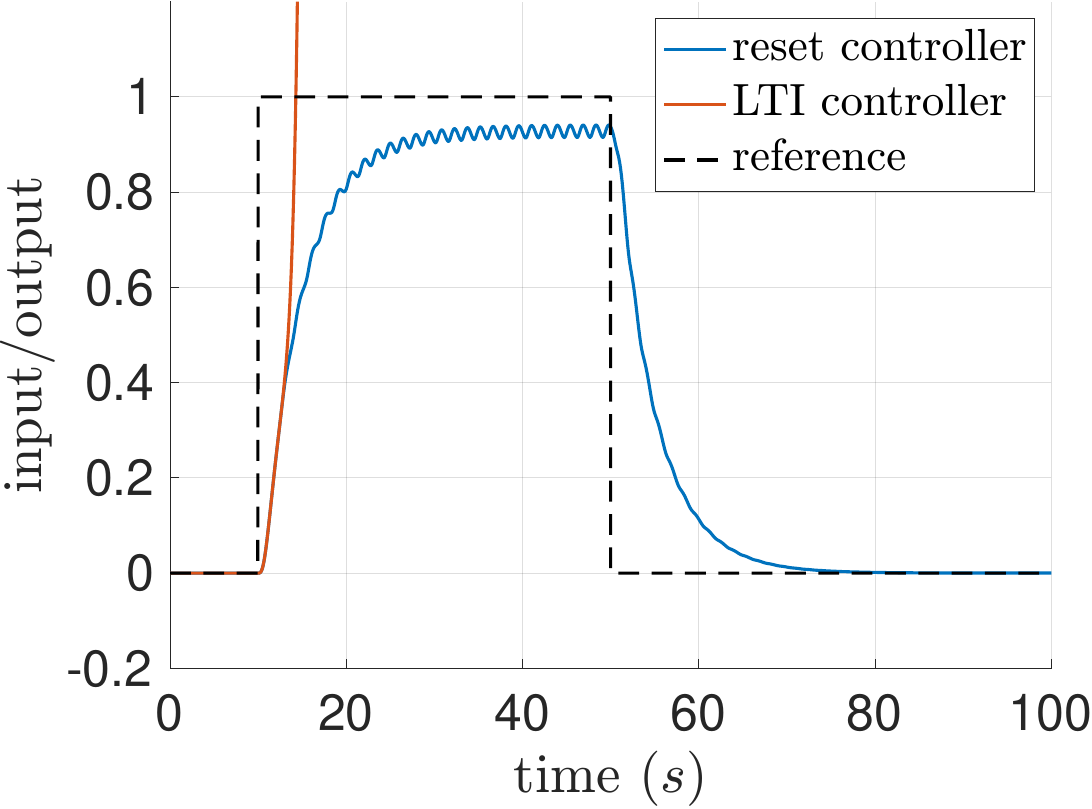}
         \caption{Simulation results.}
         \label{fig:sim_example_procedure}
     \end{subfigure}
    \caption{Analysis of Example~\ref{example:procedure}.}
    \label{fig:example_procedure}
\end{figure}

\vspace{\blocklen}
\begin{example}\label{example:procedure}
    Consider the unstable plant $G_\mathrm{u}$ from Example~\ref{example:unstable}, and consider the controller $C(k_\mathrm{p}, k_\mathrm{r}) = k_\mathrm{p}+ k_\mathrm{r} \mathcal{R}$, where $\mathcal{R}$ is defined in Example~\ref{example:reset-controller}. For $k_\mathrm{r}=-1$, find a gain value $k_\mathrm{p}$, such that $\gamma(T) \leq 1$ (i.e. $\hat{\gamma}=1$ in Eq.~\eqref{eq:reset_design_dist}), where $T = (G_\mathrm{u}^{-1} + C(k_\mathrm{p}, -1))^{-1}$.

    The bound in Eq.~\ref{eq:sg_bound_param_controller} for $k_\mathrm{p}=2.35$ is plotted in Fig.~\ref{fig:srg_example_procedure}, where we denote $\tilde{C}:= C(2.35, -1)$. We can see from Fig.~\ref{fig:srg_example_procedure} that $\operatorname{dist}(\SRG'(G)^{-1}, -2.35 + \mathcal{S}) \geq 1$ holds. In fact, $k_\mathrm{p} = 2.35$ is the smallest positive value that satisfies Eq.~\eqref{eq:reset_design_dist}.

    The simulation result is shown in Fig.~\ref{fig:sim_example_procedure}. Note that the gain bound $\gamma(T) \leq 1$ does not seem very conservative compared to the size of the step response. This is likely caused by choosing $k_\mathrm{r}$ negative, such that the $r = 0.85$ part of $\mathcal{S}$ determines the smallest separation. Unlike in Examples~\ref{example:stable} and \ref{example:unstable}, the LTI controller (i.e. where the reset condition is removed), yields an unstable closed-loop system.
\end{example}

\section{Conclusion and Discussion}

In this short note, we have shown that a reset controller, traditionally used to improve performance for stable plants, can be used to stabilize unstable plants as well. This is done by applying a gain to the reset element, and placing it in parallel with a proportional gain. Moreover, we show how the SRG framework provides a graphical tool to select these gains. It remains to be investigated how the parameters of the SORE itself can be tuned to stabilize an unstable plant. 

Since $L_2$-gain performance does not describe everything about a system, our graphical design method can be improved further by including frequency-domain method. One possible future direction is to apply recent non-approximative frequency-domain methods for SRGs~\cite{krebbekxNonlinearBandwidthBode2025} to analyze the performance of reset controllers in the frequency domain.

\begin{ack}                               
The authors would like to thank Luuk Spin and Sebastiaan van den Eijnden for insightful discussions about reset controllers. 
\end{ack}



\bibliographystyle{plain} 
\bibliography{bibliography}

\end{document}